\documentclass{article}
\usepackage{spconf,amsmath,graphicx}
\usepackage{kotex}
\usepackage{siunitx}
\usepackage{multirow}
\usepackage{amssymb}
\usepackage{pifont}
\usepackage{setspace}

\title{iPhonMatchNet: Zero-Shot User-Defined Keyword Spotting Using Implicit Acoustic Echo Cancellation}
\name{Yong-Hyeok~Lee, Namhyun~Cho}
\address{Speech AI Lab., NCSOFT Corporation, South Korea}

\begin{document}
%
\maketitle
\begin{abstract}
In response to the increasing interest in human--machine communication across various domains, this paper introduces a novel approach called iPhonMatchNet, which addresses the challenge of barge-in scenarios, wherein user speech overlaps with device playback audio, thereby creating a self-referencing problem. The proposed model leverages implicit acoustic echo cancellation (iAEC) techniques to increase the efficiency of user-defined keyword spotting models, achieving a remarkable 95\% reduction in mean absolute error with a minimal increase in model size (0.13\%) compared to the baseline model, PhonMatchNet. We also present an efficient model structure and demonstrate its capability to learn iAEC functionality without requiring a clean signal. The findings of our study indicate that the proposed model achieves competitive performance in real-world deployment conditions of smart devices.

\end{abstract}

\section{Introduction}
\label{sec:intro}


In the past few years, there has been a growing interest in various fields such as smart speakers, AI assistants, digital human interfaces, and service robots. Consequently, there is an increasing demand for personalized technologies to facilitate seamless human--machine communication. One prominent technology in this regard is user-defined keyword spotting (UDKWS), particularly in barge-in scenarios.

The term ``barge-in scenario'' refers to the phenomenon where user speech and device playback audio overlap \cite{takeda2009ica, nishimuta2014development}. This overlap not only affects keyword spotting (KWS) systems but also has critical implications for other recognition systems such as automatic speech recognition (ASR) and device-directed detection owing to the proximity of the microphone to the device's speaker rather than to the user's location \cite{huang2019study, norouzian2019exploring}. In KWS systems in particular, it can lead to the ``self-referencing'' problem, where the device mistakenly triggers itself.


A direct solution to the barge-in problem involves the use of the acoustic echo cancellation (AEC) technology. AEC estimates the room impulse response of the known playback signal, thereby removing the playback signal. Both signal processing--based research, including adaptive and Kalman filters, and neural-based research have been pursued in this field \cite{yasukawa1987acoustic, paleologu2013study, zhang2019deep, westhausen21_dtln_aec, yang2023low}.


Recent AEC challenges \cite{sridhar2021icassp, cutler2022icassp} have shown that AEC models improve not only AEC but also ASR performance. 
This AEC is typically implemented by taking the audio-out signal as a reference signal and embedding the working procedure into hardware components such as the microphone module. However, such an application is effective only in devices equipped with the specific hardware component. In other words, to provide consistent high-quality AEC technology in applications such as voice assistants across various personal devices already in use, AEC application at the software level is necessary.
However, this application requires substantial computational power, which can be a significant burden in lightweight KWS tasks. Hence, recent research has been focused on integrating implicit AEC (iAEC) techniques into KWS models \cite{cornell2023implicit}.


UDKWS is a technology that enables users to freely customize their keywords. Previous research predominantly relied on the query-by-example approach, comparing enrolled audio samples with input audio samples \cite{chen2015query, lugosch2018donut, huang2021query}. However, owing to the inconvenience of the enrollment process, recent efforts have been directed toward zero-shot UDKWS research, where keywords can be changed without requiring additional training \cite{shin22_interspeech, lee23d_interspeech, nishu2023flexible}. The performance of these models has shown remarkable improvement in recent years. Nevertheless, owing to the limitations of models that only accept microphone inputs, performance degradation occurs in situations involving human--device speech overlap.


To address these challenges, we propose iPhonMatchNet integrated with iAEC in PhonMatchNet \cite{lee23d_interspeech}. We have expanded upon our previous PhonMatchNet model to efficiently handle device playback signals and reduce false alarms in self-referencing scenarios. Through this approach, we were able to achieve an approximately 95\% reduction in mean absolute error (MAE) with only a 0.13\% increase in model size.


Our contributions can be summarized as follows:


\begin{itemize}
    \item We present a highly efficient model structure that effectively addresses the self-referencing problem.
\end{itemize}
\begin{figure*}[ht!]
    \centering
    \includegraphics[width=2.00\columnwidth]{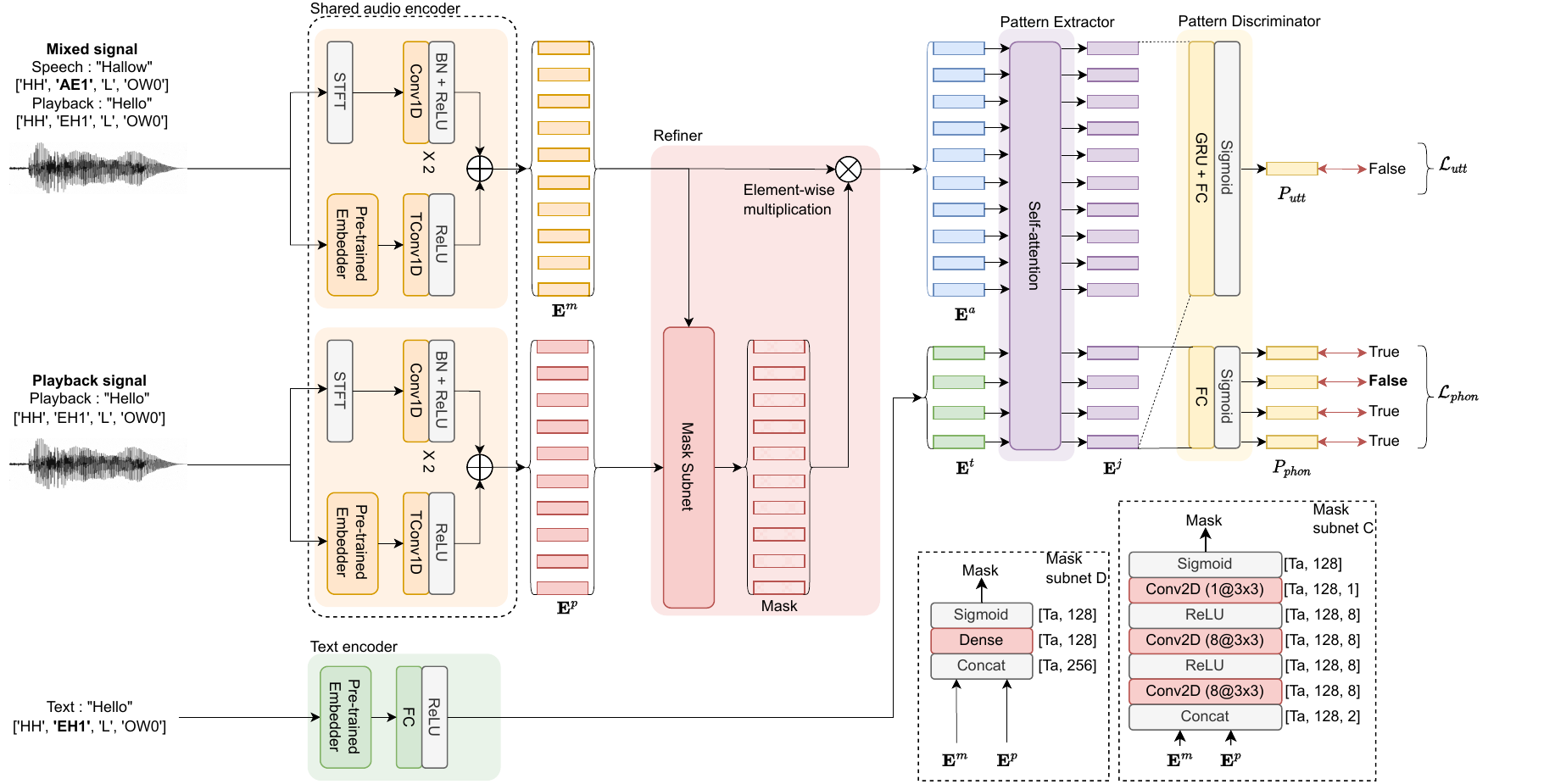}
    \caption{Architecture of the proposed model. 2D convolution layers (Conv2D) are represented as (\textit{channels}@\textit{filters}). Within the refiner, the ``Mask Subnet'' is compared in two configurations: one composed of a dense layer, referred to as ``Mask Subnet D'', and the other comprised of a convolutional layer, referred to as ``Mask Subnet C''.}
    \label{fig:overview}
\end{figure*}
\begin{itemize}
    \item We demonstrate that our model can learn iAEC functionality without the need for a clean target signal.
    \item To ensure the precise reproducibility of our experiments, we provide the code used to generate the databases adopted in our training and testing phases.
\end{itemize}

\section{Proposed method}
\label{sec:proposed}

In this section, we describe our proposed methodology, iPhonMatchNet, as illustrated in Fig. \ref{fig:overview}. The model consists of five modules - a shared audio encoder, text encoder, refiner, pattern extractor, and pattern discriminator. The model takes both mixed and playback signals, along with keyword text, as inputs to calculate whether the mixed signal, with the playback signal removed, matches the provided text. With the exception of the ``Refiner'', all components of the model structure share the same hyperparameters as in \cite{lee23d_interspeech}.

\subsection{Encoders} 

\noindent\textbf{Shared Audio Encoder.} To ensure consistency in the embedding spaces of the mixed signal and playback signal, a shared audio encoder is utilized. The audio encoder employs two feature extractors: a pre-trained speech embedder \cite{9053193} for general pronunciations and a fully trainable extractor for special pronunciations. We upsample using 1-D transposed convolution and compute 128-dimensional feature vectors. We denote playback audio embeddings as $\mathbf{E}^m, \mathbf{E}^p \in \mathbb{R}^{T_a\times128}$, where $T_a$ and 128 are the lengths of the audio and embedding dimension, respectively.

\noindent\textbf{Text Encoder.} The text encoder, similar to \cite{shin22_interspeech}, includes a pre-trained grapheme-to-phoneme (G2P) model followed by a fully connected layer and a ReLU activation function. We extract the G2P embedding from the last hidden states of the encoder. We denote text embeddings by $\mathbf{E}^t \in \mathbb{R}^{T_t\times128}$, similar to the playback audio embeddings.

\subsection{Refiner}



The refiner utilizes both $\mathbf{E}^m$ and $\mathbf{E}^p$ to create a mask, which is then applied to $\mathbf{E}^m$ using element-wise multiplication to produce the enhanced embedding $\mathbf{E}^a \in \mathbb{R}^{T_a\times128}$. To eliminate the playback-signal component from the mixed signal, which combines user speech and device playback signal, \cite{cornell2023implicit} concatenated these two signals along the embedding axis and employed dense layers with sigmoid activation to generate a mask, referred to as ``Mask Subnet D.'' However, this approach has limitations as it cannot account for the time delay resulting from the device playback signal being emitted from the speaker and re-entering the microphone input. 

Therefore, we propose ``Mask Subnet C,'' which incorporates convolutional layers to capture distortions over time. Our proposed method has been shown to outperform the approach in \cite{cornell2023implicit} with significantly fewer parameters. 

\subsection{Pattern extractor}

The pattern extractor calculates self-attention outputs using a scaled dot-product. Uni-modal embeddings, denoted as $\mathbf{E}^a$ and $\mathbf{E}^t$, are concatenated along the time dimension to create joint embeddings $\mathbf{E}^{j}$, which are then computed using the self-attention mechanism. To preserve causal information within each modality, we employ a lower triangular matrix as an attention mask.

\subsection{Pattern discriminator}

Our pattern discriminator computes two probabilities: one for audio-keyword matching and the other for audio-phoneme matching. To determine utterance-level matching, we employ a single 128-dimensional GRU layer that processes joint embeddings $\mathbf{E}^{j}$ over time. The output of the last hidden state is then passed through a fully connected layer with a sigmoid function, yielding $P_{utt}$. For phoneme-level matching, we isolate phoneme sequences from $\mathbf{E}^{j}$ and pass them through a fully connected layer with a sigmoid function, producing $P_{phon}$.

\subsection{Training criterion}

We employed the same training criterion as in \cite{lee23d_interspeech}. For all utterance-level and phoneme-level detection losses, we utilized the ground truth from the labels of the target speech data within the mixed signal.

\section{Experiments}
\label{sec:experiments}

Here, we describe the experimental setup, including the datasets, evaluation metrics, and implementation details for training and testing. 

\subsection{Datasets}
\label{sec:datasets}

We used four datasets for training and evaluation: LibriPhrase \cite{shin22_interspeech}, Google Speech Commands V1 ($\mathbf{G}$) \cite{warden2018speech}, Qualcomm Keyword Speech dataset ($\mathbf{Q}$) \cite{kim2019query}, and MUSAN \cite{musan2015}. The MUSAN dataset was utilized as the playback-signal source, while the remaining datasets were employed for the UDKWS. In the training phase, we used the training set of LibriPhrase and babble noise from the MS-SNSD dataset \cite{reddy2019scalable} for robust detection. During evaluation, we used the datasets $\mathbf{G}$, $\mathbf{Q}$, LibriPhrase-easy ($\mathbf{LP_E}$), and LibriPhrase-hard ($\mathbf{LP_H}$), detailed in \cite{lee23d_interspeech}. To enhance iAEC performance, we generated data for both the training and evaluation phases by incorporating music and speech data from MUSAN as playback signals. Detailed procedures are outlined in Section \ref{sec:rir}.

\subsection{Room acoustics simulation}
\label{sec:rir}

To simulate barge-in scenarios involving device playback signals, we synthesized mixed signals by convolving target speech data with room impulse responses (RIR). We considered four distinct cases based on the type of playback signal: 1) non-playback, 2) playback - music, 3) playback - speech, and 4) self-referencing. 
The ``non-playback'' scenario occurs when there is user speech but no playback signal present. In the ``playback - music'' and ``playback - speech'' cases, the device playback signal exists. During these scenarios, the playback signal types are divided into music and speech, and they are synthesized accordingly using MUSAN data. The ``self-referencing'' case represents a significant area of focus, where the device playback signal matches the target speech signal. This scenario occurs when keywords such as ``Hey Siri'' or ``Hi Bixby'' are generated by the smartphone itself.

To accommodate these scenarios, we positioned audio sources accordingly and performed room acoustics simulations using pyroomacoustics \cite{scheibler2018pyroomacoustics} under the following random conditions: room size sampled from a uniform distribution $U(10,50)$ $m^2$, room height from $U(2.5,5.0)$ $m $, reverberation time from $U(0.2,0.6)$ $s$, propagation delay from $U(0.01,0.1)$ $s$, and signal-to-interference ratio from $U(-12,3)$ dB. We have made the scripts for generating this dataset publicly available\footnote{https://github.com/ncsoft/PhonMatchNet}.

Furthermore, to account for real device recording conditions, we exclusively utilized the output of the simulation, i.e., the mixed signal. In most edge devices, there is no access to the user's clean speech, and they can only record the sound from their speakers and microphone inputs. Therefore, even though we had access to clean speech data, we refrained from using these during training and solely relied on the mixed-signal data.

\subsection{Metrics}

We evaluated our proposed model by measuring the equal error rate (EER), area under the curve (AUC), and mean absolute error (MAE) at the sample level on these datasets. In the ``self-referencing'' case, since all ground truth values are 0, we evaluated the model's performance using only MAE because EER and AUC were not applicable. For the other scenarios, we assessed the model using EER and AUC.

\subsection{Implementation details}

The training criterion was optimized using the Adam optimizer with the default parameters. The models were trained for 100 epochs with a fixed learning rate of $10^{-3}$, and the best model was selected based on performance on the test sets. For training, we used a single V100 with a batch size of 1024 for a week.

\begin{table*}[ht!]
\centering
\caption{Performance of the baseline, proposed method, and ablations on various datasets. $\mathbf{G}$, $\mathbf{Q}$, $\mathbf{LP_E}$, and $\mathbf{LP_H}$ refer to our test sets from Section \ref{sec:datasets}. NLMS denotes Normalized Least Mean Square, based on adaptive filter, an algorithm employed in acoustic echo cancellation (AEC). The filter length and step size parameters are the same as those in \cite{cornell2023implicit}.}
\label{tab:my-table}
\resizebox{2.0\columnwidth}{!}{%
\begin{tabular}{c|c|c|c|cc|cc|cc|c}
\hline
\multirow{2}{*}{Model} &
  \multirow{2}{*}{AEC} &
  \multirow{2}{*}{Params} &
  \multirow{2}{*}{Dataset} &
  \multicolumn{2}{c|}{Non-playback} &
  \multicolumn{2}{c|}{Music} &
  \multicolumn{2}{c|}{Speech} &
  Self-referencing \\ \cline{5-11} 
 &
   &
   &
   &
  \multicolumn{1}{c|}{AUC (\%) $\uparrow$} &
  EER (\%) $\downarrow$ &
  \multicolumn{1}{c|}{AUC} &
  EER &
  \multicolumn{1}{c|}{AUC} &
  EER &
  MAE (\%) $\downarrow$ \\ \hline
\multirow{8}{*}{\begin{tabular}[c]{@{}c@{}}PhonMatchNet\\ \cite{lee23d_interspeech}\end{tabular}} &
  \multirow{4}{*}{-} &
  \multirow{8}{*}{655 K} &
  $\mathbf{G}$ &
  \multicolumn{1}{c|}{\textbf{98.11}} &
  6.77 &
  \multicolumn{1}{c|}{88.85} &
  19.39 &
  \multicolumn{1}{c|}{71.61} &
  33.66 &
  73.97 \\
 &
   &
   &
  $\mathbf{Q}$ &
  \multicolumn{1}{c|}{98.90} &
  4.75 &
  \multicolumn{1}{c|}{82.37} &
  26.13 &
  \multicolumn{1}{c|}{71.47} &
  33.76 &
  71.72 \\
 &
   &
   &
  $\mathbf{LP_E}$ &
  \multicolumn{1}{c|}{99.29} &
  2.80 &
  \multicolumn{1}{c|}{89.31} &
  19.40 &
  \multicolumn{1}{c|}{79.52} &
  27.17 &
  82.75 \\
 &
   &
   &
  $\mathbf{LP_H}$ &
  \multicolumn{1}{c|}{88.52} &
  \textbf{18.82} &
  \multicolumn{1}{c|}{71.21} &
  34.94 &
  \multicolumn{1}{c|}{63.51} &
  40.35 &
  85.40 \\ \cline{2-2} \cline{4-11} 
 &
  \multirow{4}{*}{NLMS} &
   &
  $\mathbf{G}$ &
  \multicolumn{1}{c|}{98.08} &
  \textbf{6.73} &
  \multicolumn{1}{c|}{87.64} &
  20.83 &
  \multicolumn{1}{c|}{73.24} &
  31.78 &
  72.55 \\
 &
   &
   &
  $\mathbf{Q}$ &
  \multicolumn{1}{c|}{\textbf{99.29}} &
  \textbf{3.90} &
  \multicolumn{1}{c|}{82.50} &
  26.52 &
  \multicolumn{1}{c|}{72.14} &
  32.46 &
  69.86 \\
 &
   &
   &
  $\mathbf{LP_E}$ &
  \multicolumn{1}{c|}{\textbf{99.59}} &
  2.70 &
  \multicolumn{1}{c|}{89.78} &
  19.28 &
  \multicolumn{1}{c|}{80.06} &
  27.34 &
  72.91 \\
 &
   &
   &
  $\mathbf{LP_H}$ &
  \multicolumn{1}{c|}{\textbf{88.71}} &
  19.45 &
  \multicolumn{1}{c|}{71.98} &
  34.47 &
  \multicolumn{1}{c|}{63.81} &
  40.01 &
  72.12 \\ \hline
\multirow{8}{*}{iPhonMatchNet} &
  \multirow{4}{*}{\begin{tabular}[c]{@{}c@{}}iAEC\\ SubnetMask D\end{tabular}} &
  \multirow{4}{*}{688 K} &
  $\mathbf{G}$ &
  \multicolumn{1}{c|}{96.30} &
  9.89 &
  \multicolumn{1}{c|}{88.64} &
  20.20 &
  \multicolumn{1}{c|}{77.37} &
  29.97 &
  0.70 \\
 &
   &
   &
  $\mathbf{Q}$ &
  \multicolumn{1}{c|}{98.30} &
  6.06 &
  \multicolumn{1}{c|}{83.41} &
  25.66 &
  \multicolumn{1}{c|}{76.45} &
  30.71 &
  0.65 \\
 &
   &
   &
  $\mathbf{LP_E}$ &
  \multicolumn{1}{c|}{99.33} &
  3.30 &
  \multicolumn{1}{c|}{93.62} &
  14.61 &
  \multicolumn{1}{c|}{88.49} &
  19.98 &
  1.44 \\
 &
   &
   &
  $\mathbf{LP_H}$ &
  \multicolumn{1}{c|}{85.88} &
  21.74 &
  \multicolumn{1}{c|}{74.51} &
  32.45 &
  \multicolumn{1}{c|}{68.90} &
  36.63 &
  7.74 \\ \cline{2-11} 
 &
  \multirow{4}{*}{\begin{tabular}[c]{@{}c@{}}iAEC\\ SubnetMask C\end{tabular}} &
  \multirow{4}{*}{\textbf{656 K}} &
  $\mathbf{G}$ &
  \multicolumn{1}{c|}{96.19} &
  8.09 &
  \multicolumn{1}{c|}{\textbf{92.03}} &
  \textbf{15.22} &
  \multicolumn{1}{c|}{\textbf{80.35}} &
  \textbf{26.63} &
  \textbf{0.15} \\
 &
   &
   &
  $\mathbf{Q}$ &
  \multicolumn{1}{c|}{98.90} &
  5.67 &
  \multicolumn{1}{c|}{\textbf{85.62}} &
  \textbf{23.05} &
  \multicolumn{1}{c|}{\textbf{78.80}} &
  \textbf{28.93} &
  \textbf{0.13} \\
 &
   &
   &
  $\mathbf{LP_E}$ &
  \multicolumn{1}{c|}{\textbf{99.59}} &
  \textbf{2.40} &
  \multicolumn{1}{c|}{\textbf{95.12}} &
  \textbf{12.76} &
  \multicolumn{1}{c|}{\textbf{89.62}} &
  \textbf{18.57} &
  \textbf{0.31} \\
 &
   &
   &
  $\mathbf{LP_H}$ &
  \multicolumn{1}{c|}{88.23} &
  19.70 &
  \multicolumn{1}{c|}{\textbf{77.40}} &
  \textbf{30.36} &
  \multicolumn{1}{c|}{\textbf{72.01}} &
  \textbf{32.79} &
  \textbf{3.11} \\ \hline
\end{tabular}%
}
\end{table*}

\section{Results}
\label{sec:results}

\subsection{Comparison with baseline}

Our proposed model outperformed the baseline model (PhonMatchNet), which does not take the playback signal as input, in all cases except for the non-playback scenario. Even in non-playback scenarios, our model exhibited only a slight increase in error relative to the baseline model. Particularly, in the self-referencing scenario, which is the focus of our concern, our approach showed a dramatic reduction in error rate. In this scenario, where all ground truth values are 0, the MAE value is equal to the average output of the model. Therefore, we can conclude that the model has a low probability of mistakenly responding in most of the test cases. This confirms the suitability of our proposed method for barge-in scenarios. 

\subsection{Comparison between Mask Subnet D and C}

Among our proposed methods, ``Mask Subnet D'' and ``Mask Subnet C'' exhibit a significant difference in parameter size, yet their performance shows the opposite trend. This is attributed to the relationship between the mixed and playback signals, favoring convolutional over linear layers. Because playback signals inevitably introduce time delays when re-entering the microphone, ``Mask Subnet C'', which can account for this, proves to be a more efficient masking approach.

\subsection{Comparison with conventional AEC method}

We experimented by applying the widely adopted NLMS method of AEC to the existing baseline model. The results showed that there was no significant performance improvement in ``Music'' and ``Speech'' playback scenarios. Instead, we observed performance enhancement in the ``Non-playback'' case. However, in ``Self-referencing'' scenarios, there was no significant performance enhancement. This is interpreted as the influence of residual echo that was not filtered out by AEC, which the model detected. Improving the robustness of the existing UDKWS model with AEC may lead to even worse performance in ``self-referencing'' situations. Therefore, it can be concluded that our proposed model, which includes the iAEC structure, is an effective solution.

\section{Conclusions}
\label{sec:conclusions}

This study introduces a novel zero-shot user-defined keyword spotting model suitable for barge-in scenarios. Our proposed model is designed to receive both microphone input and playback signals, enabling it to ignore signals played back by the device. We demonstrated that this architecture led to improved performance in various settings compared with existing models, particularly addressing the significant vulnerability to self-referencing issues in KWS systems. Furthermore, we achieved competitive performance without utilizing clean user speech during training, reflecting the real-world deployment conditions of smart devices.
Despite our extensive efforts, we observed relatively lower performance improvements in challenging cases, such as ``playback - speech,'' compared to other test sets. Our future research direction is to enhance the performance in highly correlated playback cases.

\begin{spacing}{0.81}
\bibliographystyle{IEEEbib}
\bibliography{strings,refs}
\end{spacing}

\end{document}